\documentstyle[dina4,12pt]{article}

\begin{document}

\newcommand{\be}{\begin{equation}}
\newcommand{\ee}{\end{equation}}
\newcommand{\bea}{\begin{eqnarray}}
\newcommand{\eea}{\end{eqnarray}}
\newcommand{\no}{\nonumber \\}
\newcommand{\fs}{\; \; .}
\newcommand{\co}{\; \; ,}
\newcommand{\eff}{e\hspace{-0.1em}f\hspace{-0.18em}f}

\begin{titlepage}
\begin{flushright}BUTP-93/25\end{flushright}
\rule{0em}{2em}\vspace{2em}
\begin{center}
{\LARGE {\bf Nonrelativistic effective Lagrangians}}\\ \vspace{2em}
H. Leutwyler\\Institut f\"{u}r theoretische Physik der Universit\"{a}t
Bern\\Sidlerstr. 5, CH-3012 Bern, Switzerland\\
\vspace{2em}
September 1993\\
\vspace{3em}
{\bf Abstract} \\
\vspace{2em}
\parbox{30em}{Chiral perturbation theory is extended to nonrelativistic
systems with spontaneously broken symmetry.
In the effective Lagrangian, order parameters associated with the
generators of the group manifest themselves as effective coupling
constants of a topological term, which is gauge invariant only up to a total
derivative. In the case of the
ferromagnet, a term connected with the Brouwer degree dominates the derivative
expansion. The general analysis
includes antiferromagnetic magnons and phonons, while the
effective field theory of
fluids or gases is beyond the scope of the method.}\\
\vspace{1em}
\rule{30em}{.02em}\\
{\footnotesize Work
supported in part by Schweizerischer Nationalfonds}
\end{center}
\end{titlepage}

\section{Introduction}
\label{Introduction}

The various low energy phenomena considered in the present paper are very well
explored, at a level which goes much beyond the general discussion given below.
The aim of the paper is not to contribute to the detailed physical
understanding of the many different systems known to exhibit spontaneous
symmetry breakdown, but
to analyze their low energy structure from a unified
point of view, relying on the method of effective Lagrangians. This
method
is widely used in condensed matter physics \cite{Anderson}, but, as far as I
know, a general analysis is
not available. In particular, an effective
Lagrangian describing the behaviour of a ferromagnet at large
wavelengths does not appear to
exist in the literature.
The main result of the present work is an expression for the general effective
Lagrangian. As it turns out, the expression contains a term of rather
remarkable structure, which distinguishes ferromagnets from
other systems.

The analysis is based on general considerations, applicable to any system,
for which the Goldstone modes represent the only excitations without an energy
gap. It amounts to an extension of the effective theories used in
particle physics
\cite{Dashen}--\cite{GL} to the nonrelativistic
domain. This extension is by no means
trivial. The
relativistic situation is considerably simpler, because Lorentz invariance
imposes strong constraints on the low energy structure of the theory and, e.g.,
prevents
the charge densities
from picking up an expectation value in the ground state. These constraints
do not apply to condensed matter, where
the center of mass distinguishes a preferred frame of reference. Moreover,
the lattice
structure of a solid singles out preferred directions, such that the effective
Lagrangian is not invariant under rotations, either.
In the case of a cubic lattice,
the anisotropy, however, only shows up in the higher orders of the derivative
expansion. As the following discussion mainly concerns the leading
contributions, I disregard from this complication and
assume
that, at large distances, the correlation functions are invariant,
both, under translations and rotations.

I consider a spontaneously broken exact symmetry in $d=3+1$ dimensions
(spontaneous breakdown of symmetries only occurs for
$d>2$ --- the low energy behaviour of the two-dimensional
nonlinear $\sigma$-model, e.g., cannot be
analyzed in terms of an effective Lagrangian \cite{Mermin}). The
Hamiltonian is symmetric with respect to a Lie group G with generators $Q_i$,
\be\label{ferro1}
[Q_i,H]=0\hspace{2em},\hspace{2em}[Q_i,Q_j]=if^k_{\;ij}Q_k \co\hspace{2em}\ee
but the ground state
$\mid\!0\!>$ is invariant only under a subgroup $\mbox{H}\subset \mbox{G}$.
For Lorentz invariant theories, the Goldstone theorem \cite{Goldstone} states
that the spontaneous symmetry breakdown gives rise to
$\mbox{dim(G)}-\mbox{dim(H)}$ massless particles. In the non\-re\-la\-tivistic
regime, the occurrence of order parameters also implies that there are modes of
excitation, for which the frequency $\omega$
disappears when the wave vector $\vec{k}$ tends to zero, but the number
of independent such states and their dispersion law depend on the properties
of the system \cite{Anderson,Guralnik}.

The generators
$Q_i$ of G are space integrals over the corresponding charge densities,
\be\label{ferro2}
Q_i=\int\!d^3\!x J^0_i(x)
\fs\ee
Identifying the zeroth component of the coordinate vector with
the time, ($x^0=t$, no factor of $c$),
charge conservation takes the local form
\be\label{ferro3}
\partial_\mu J^\mu_i(x)\equiv\partial_0J^0_i(x)+\partial_rJ^r_i(x)=0\fs\ee

The time-ordered correlation functions of the charge densities
$J^0_i(x)$ and currents $J^r_i(x)$ play a central role
in the analysis of the low energy structure: the construction of the effective
theory relies on the Ward identities, which
express the
symmetry properties of the system in terms of these quantities.
It is convenient to collect the correlation functions in a generating
functional $\Gamma\{f\}$, \be\label{ferro4}
e^{i\,\Gamma\{f\}}=\sum_{n=0}^\infty\frac{i^n}{n!}\int\!d^4\!x_1\ldots
d^4\!x_n \;f_{\mu_1}^{i_1}(x_1)\ldots f_{\mu_n}^{i_n}(x_n)
<\!0\!\mid T\{J^{\mu_1}_{i_1}(x_1)\ldots
J^{\mu_n}_{i_n}(x_n)\}\mid\!0\!>\ee
where $\mid\!0\!>$ denotes the ground state of the system and $f_\mu^i(x)$ is
an external field, which plays the role of an auxiliary variable.
The generating functional describes the transitions which occur if
the system is perturbed by an external field, $H\rightarrow\, H-\int\!d^3x
f_\mu^iJ^\mu_i$. The quantity $e^{i\,\Gamma\{f\}}$ is the probability
amplitude for the system to remain in the ground state for
$t\rightarrow +\infty$, if it was there at
$t\rightarrow -\infty$.

If the
theory does not contain anomalies, the Ward identities are equivalent to the
statement that the generating functional is invariant under gauge
transformations of the external field,
\be\label{ferro5}
\delta f_\mu^i(x)=D_\mu g^i(x)\equiv \partial_\mu g^i(x)
+f^i_{\;jk}f_\mu^j(x)g^k(x)\hspace{2em}\rightarrow\hspace{2em}
\delta\Gamma\{f\}=0
 \fs\hspace{2em}\ee
The gauge functions $g^1(x), g^2(x),\ldots$ are arbitrary infinitesimal
quantities. They may be viewed as coordinates of a space-time dependent
group element
$g(x)\in\mbox{G}$ in the infinitesimal neighbourhood of unity.

The low energy analysis concerns the behaviour of the correlation functions at
distances large compared to the intrinsic scales of the theory. In particular,
the distances under consideration are assumed to be large compared to the
lattice spacing $a$ --- the effective theory does not resolve the
microscopic structure of the system, i.e., refers to the continuum limit.
In
the language of the generating functional, the effective theory concerns slowly
varying external fields, such that $\partial f/\partial x \ll  f/a$.

The Fourier transforms of the various correlation functions contain
singularities at low energies and momenta,
due to the propagation of Goldstone excitations. The singularities arise
from processes involving the
emission of Goldstone bosons, which travel over a long distance
before being
absorbed. In particular, one-particle-reducible
contributions
generate poles, while the simultaneous exchange
of several Goldstone modes produces cuts.

\section{Effective Lagrangian}
\label{Effective}

The following discussion exclusively deals with the contributions due to
the Goldstone excitations.
As witnessed by superconductivity
or by the Higgs sector of the Standard Model, the presence of additional
degrees of freedom without energy gap may change the low energy structure, even
qualitatively: gauge fields may eat the Goldstone bosons up. In the
following, it is assumed that, at low frequencies and large
wavelengths, the spectrum exclusively contains Goldstone excitations.
More precisely, the discussion relies on the
PCAC hypothesis, according to which the poles generated by the
Goldstone bosons dominate the low energy behaviour of the correlation
functions.

As is well-known, the singularities due to the
exchange of Goldstone bosons may be described in terms of an effective field
theory \cite{Dashen,Callan}. I refer to the variables of the effective theory
as "pion" fields, using the symbol
$\pi^a(x)$ (in the applications to be
discussed below, the "pions" represent magnons or phonons). Unlike the number
of
Goldstone {\it particles}, which depends on the form of the dispersion law, the
number of {\it fields} needed to
describe them is universal: the effective theory involves
$\mbox{dim(G)}-\mbox{dim(H)}$ real fields. If the dispersion law is of the form
$\omega(\vec{k})=v\!\!\mid\!\!\vec{k}\!\!\mid\!+\,O(k^2)$, as it is the case
for
Lorentz invariant theories, the number of independent one-particle-states of
momentum
$\vec{k}$ is the same as the number of fields. For a dispersion law of the type
$\omega(\vec{k})=\gamma\vec{k}^{\,2}+O(k^4)$, on the other hand, the
number of states is given by
$\frac{1}{2}\{\mbox{dim(G)}-\mbox{dim(H)}\}$. The difference is related to the
order of the corresponding wave equations. In the first case, the
wave equation is of second order in
the time derivatives. The Fourier decomposition then contains both, positive
and negative frequencies and a real field suffices to describe a
particle. In the second case, the wave equation takes the
form of the Schr\"{o}dinger equation, such that only positive frequencies occur
and a complex field is needed per particle.

In the language of the effective field theory, the one-particle-reducible
contributions responsible for the poles are represented by the tree graphs. The
pole terms arise from pion field propagators, whose form is specified by
the kinetic part of the effective Lagrangian, i.e., by the part which is
quadratic in the pion field. The interaction terms of the effective
Lagrangian are in one-to-one correspondence with
the amplitudes for emission, absorption and scattering.
In addition to the purely pionic vertices,
describing the interaction of the Goldstone bosons among themselves, the
Lagrangian also contains
vertices involving the external field, which describe the
transitions generated by the perturbation
$f_\mu^iJ^\mu_i$. The matrix element
$<\!0\!\mid\! f_\mu^iJ^\mu_i\!\mid\!\pi\!>$,
e.g., which represents the probability amplitude for the external field to
excite one of the Goldstone states, is represented in the effective Lagrangian
through a term linear in the fields $f_\mu^i(x),\,\pi^a(x)$.

The low
energy analysis
relies on an expansion of the vertices in powers of the momenta. In the
language of the effective field theory, this corresponds to a derivative
expansion of the Lagrangian, ${\cal L}_{\eff}={\cal L}_{\eff}(\pi,\partial_\mu
\pi,
\partial_\mu\partial_\nu \pi,\ldots;f,\partial_\mu f,\ldots )$. The generic
term occurring therein contains $P$ pion fields, $E$ external fields
and, altogether, $D$ derivatives, some acting on $\pi^a(x)$, some on
$f_\mu^i(x)$. It is convenient to count the external field on the same
footing as the derivatives, $f_\mu^i\propto\partial_\mu$, but to
distinguish between the time and space components of these quantities. The
derivative expansion then consists of a double series of the form
\be\label{ferro6}{\cal L}_{\eff}=\sum_{s,\,t}{\cal L}_{\eff}^{(s,\,t)}\fs\ee
The term ${\cal L}_{\eff}^{(0,0)}$ exclusively contains the pion field and
does not involve derivatives, ${\cal L}_{\eff}^{(0,1)}$ collects the purely
pionic vertices with one time derivative, as well as those involving one
factor of $f_0^i$, but no derivatives, etc. Note that the number of Goldstone
bosons entering
the vertices is not specified --- the various terms occurring in
the derivative expansion represent functions of the pion field.

The tree graphs yield the leading term in the low
energy expansion of the generating functional, loops only generating
corrections of nonleading order \cite{Weinberg1979}.
The tree graphs of a quantum field theory represent the corresponding classical
field theory. More precisely, the tree graph contributions to the generating
functional are given by the classical action,
\be\label{ferro7}
\Gamma\{f\}\rule[-.5em]{.04em}{1.2em}_{\;\mbox{\scriptsize tree}}
=S_{\eff}\{\pi,f\}\;\;\;\;,\;\;\;\;S_{\eff}
\{\pi,f\} \equiv\int\!d^4\!x{\cal
L}_{\eff}(\pi,\partial\pi,\ldots;f,\partial f,\ldots)\fs\ee
The action is to be evaluated at the extremum, where the pion field obeys the
classical equation of motion
\be\label{ferro8}
\frac{\delta S_{\eff}\{\pi,f\}}{\delta \pi^a(x)}=0\fs\ee

The Ward
identities are obeyed if and only if the generating functional is gauge
invariant. For this to be the case at leading order of the low energy
expansion, the value of the classical action at the extremum must
be gauge invariant, \be\label{ferro9}
D_\mu\frac{\delta S_{\eff}\{\pi,f\}}{\delta f_\mu^i(x)}=0\fs\ee

The pion field thus simultaneously obeys the two differential equations
(\ref{ferro8}) and (\ref{ferro9}). While the first one is the standard
equation of motion,
the second incorporates the Ward identities connected with the hidden symmetry
and very strongly constrains the form of the Lagrangian. In fact,
this constraint determines the leading terms of the derivative
expansion up to a few constants, which play the role of effective
coupling constants (a detailed analysis of the same two differential
equations for the case of a Lorentz invariant effective theory
is given in ref. \cite{Foundations}).

\section{Leading Orders of Derivative Expansion}
\label{Leading}

For the framework to be internally consistent, the form of the two differential
equations (\ref{ferro8}) and (\ref{ferro9}) must be compatible with
the derivative expansion.
The leading term occurring in that expansion, ${\cal
L}_{\eff}^{(0,0)}$, does not contain derivatives of the pion field. To leading
order,
the "equation of motion" then reduces to a purely algebraic
condition on this field, $\partial {\cal L}_{\eff}^{(0,0)}(\pi)/\partial
\pi^a=0$. It is evident that the loop expansion does not make sense if the
kinetic term
only occurs among the higher order corrections:
if this were so, the pions would not propagate at all, the "propagator" taking
the form of a $\delta$-function. Indeed, it is well-known that the hidden
symmetry not only protects the Goldstone bosons from acquiring mass, but also
suppresses their mutual interactions at low energies. Current conservation
implies that all of the
vertices disappear if the momenta become small, such that
purely pionic vertices without derivatives
do not occur, ${\cal L}_{\eff}^{(0,0)}=0$.

The derivative expansion of the effective Lagrangian
thus starts with ${\cal L}_{\eff}^{(0,1)}$. Invariance under space rotations
permits two contributions of this order:\footnote{Notation:
$i,j,k =1,\ldots,\mbox{dim(G)}$
label the generators of the group,
$a,b,c=1,\ldots,\mbox{dim(G)}\!-\!\mbox{dim(H)}$ denote the components
of the effective field and $r,s,t=1,2,3$ refer to the spacial
coordinates. Repeated indices are summed over.}
\be\label{ferro10}
{\cal L}_{\eff}^{(0,1)}= c_a(\pi)\dot{\pi}^a+ e_i(\pi)f^i_0\fs\ee
The space derivatives of the pion field only show up at the next order of the
expansion, where the general form of the Lagrangian
consistent with rotation symmetry reads
\bea\label{ferro11}
{\cal L}_{\eff}^{(2,0)}&\hspace{-0.5em}=&\hspace{-0.5em}
\,-\mbox{$\frac{1}{2}$}g_{ab}(\pi )\,\partial_r \pi^a \partial_r\pi^b
+ h_{ai}(\pi )f^i_r\partial_r \pi^a
-\mbox{$\frac{1}{2}$}k_{ik}(\pi)f^i_s f^k_s  \no
{\cal L}_{\eff}^{(0,2)}&\hspace{-0.5em}=&\hspace{-0.5em}
\;\mbox{$\frac{1}{2}$}\bar{g}_{ab}(\pi )\,
\dot{\pi}^a \dot{\pi}^b
 - \bar{h}_{ai}(\pi )f^i_0\;\dot{\pi}^a
+\mbox{$\frac{1}{2}$}\bar{k}_{ik}(\pi)f^i_0 f^k_0\fs
\eea
Note that terms involving second derivatives of the pion field or
first derivatives of the external field may be removed by adding a suitable
term of the form $\partial_\mu\omega^\mu$, which does not contribute to the
action.

The term ${\cal L}_{\eff}^{(0,1)}$ does not occur in Lorentz invariant
effective theories --- it represents the main novelty in the
extension of these to nonrelativistic systems (in addition, Lorentz
invariance implies that the functions
$\bar{g}_{ab}(\pi),\bar{h}_{ai}(\pi),\bar{k}_{ik}(\pi)$ coincide with the
corresponding unbarred quantities, up to a factor of $c^2$).
The value of $e_i(\pi)$ at $\pi=0$ yields a term
in the effective Lagrangian, which is linear in the external field and
hence determines the one-point-function,
\be\label{ferro12}
<\!0\!\mid \!J^0_i(x)\!\mid\!0\!>=e_i(0)\fs\ee

For nonabelian symmetries, the charge densities transform in a
nontrivial manner under G, such that their expectation values represent
order parameters.
The ground state of a {\it ferromagnet}, e.g., singles out a direction of the
magnetization, given by the expectation value of the spin density. The
corresponding "charges" generate the group G=O(3) of spin rotations.
The direction of the magnetization
need not be correlated
with the orientation of the lattice. For the Heisenberg model, e.g.,
the spin rotations play the role of an internal symmetry; at
long wavelengths, the Green functions of this model are indeed invariant under
euclidean transformations of three-dimensional space, as it is assumed
here.
The case of the {\it antiferromagnet} shows, however,
that the expectation values of the charge densities are not necessarily
different
from zero. The constants $e_i(0)$ represent coupling constants of the effective
Lagrangian; symmetry alone does not tell what values these constants take.

\section{Symmetry properties of the Lagrangian}
\label{Symmetry}

As discussed above, the pion field must simultaneously obey the equation of
motion (\ref{ferro8}) and the Ward identity (\ref{ferro9}).
In general,
the leading term in the derivative expansion of the equation of motion is of
first order in the time derivative, while the space derivatives only enter at
second order, through a term from ${\cal L}_{\eff}^{(2,0)}$, proportional to
$\bigtriangleup \pi$. The equation of
motion thus takes the form of a Schr\"{o}dinger equation, leading to a
dispersion law of the type $\omega\propto \vec{k}^2$. It is convenient to
organize the bookkeeping accordingly,
counting energies like two powers of momenta. The terms
${\cal L}_{\eff}^{(0,1)}$ and
${\cal L}_{\eff}^{(2,0)}$ then represent expressions of the same order $k^2$,
while the remainder of the derivative expansion is of order $k^3$ or higher.
The Ward identity is of the same form as the equation of motion, also
relating $\dot{\pi}$ to $\bigtriangleup \pi$. The two equations are
consistent with one another only if they are linearly dependent.
Solving
the equation of motion for $\dot{\pi}$ and inserting the result in
the Ward identity, one obtains a relation which only
involves the pion field, its spacial derivatives and the external field. Since
these are independent from one another, the condition is obeyed only if the
coefficients occurring therein are equal to zero. This subjects
the functions
$c_a(\pi),e_i(\pi),g_{ab}(\pi),h_{ai}(\pi),k_{ik}(\pi)$,
which specify the vertices of the effective Lagrangian,
to the following conditions:
\bea\label{ferro13}
&(a)&\;\;d_ih^a_{\;j}-d_jh^a_{\;i}= f^k_{\;ij}h^a_{\;k}\co\no
&(b)&\;\;\nabla_{\!a}h_{bi}+\nabla_{\!b}h_{ai}=0\co\hspace{6em}\no
&(c)&\;\;k_{ik}=g^{ab}h_{ai}h_{bk}\co\no
&(d)&\;\;d_ie_j= f^k_{\;ij}e_k\co\no
&(e)&\;\;h^b_{\;i}\,(\partial_bc_a-\partial_ac_b)=\partial_ae_i\fs\eea
To simplify these formulae, I have used the following notation:
The matrix $g_{ab}(\pi)$ plays the
role of a metric on the manifold of pion field variables. Indices are lowered
and raised with this metric and its inverse, $g^{ab}(\pi)$, e.g.,
$h^a_{\;i}= g^{ab}h_{bi}$. The symbol $\nabla_{\!a}$ is the
corresponding covariant derivative, formed with the Christoffel symbol,
$\nabla_{\!a}h_{bi}= \partial_ah_{bi}-\Gamma^c_{ab}h_{ci}$ and
$d_i$ stands for the differential operator
$d_i= h^a_{\;i}(\pi)\partial_a$. Note that the above relations exclusively
involve
derivatives with respect to the pion field variables, which represent the
arguments of the functions occurring in the effective Lagrangian,
$\partial_a\equiv\partial/\partial\pi^a$.

The first three relations are identical with those relevant in the relativistic
case, where $c_a(\pi)=e_i(\pi)=0$. They state that
the metric $g_{ab}(\pi)$ describes a symmetric space with isometry group G.
The functions $h^a_{\;i}(\pi)$ represent the corresponding Killing
vectors, which specify the shift in the pion field generated by
infinitesimal group motions,
\be\label{ferro14} \delta\pi^a=h^a_{\;i}(\pi)\,g^i\fs\ee
The geometry of the groups G and H fixes the functions $h^a_{\;i}(\pi)$, except
for the choice of field variables. The symmetry also very strongly constrains
the form of the metric. In particular, if the Goldstone bosons transform
irreducibly under H, the metric is fixed up to an
effective coupling constant $F$: denoting the intrinsic metric of the
quotient space G/H by
$\hat{g}_{ab}(\pi)$, the metric relevant for the effective Lagrangian is given
by $g_{ab}(\pi)=F^2\,\hat{g}_{ab}(\pi)$. A detailed discussion
of these statements is given in ref. \cite{Foundations}, where
it is also shown that the conditions $(a)$, $(b)$ and $(c)$ insure invariance
of ${\cal L}_{\eff}^{(2,0)}$ under a simultaneous gauge transformation of
the fields $f_\mu^i(x)$ and $\pi^a(x)$. For the case of an abelian symmetry,
the coordinates may be chosen such that, both the Killing vectors and the
metric are constants, $h^a_{\;i}(\pi)=h^a_{\;i}(0),\,g_{ab}(\pi)=g_{ab}(0)$.

The new couplings
$e_i(\pi)$ and $c_a(\pi)$ only occur in the conditions
$(d)$ and $(e)$. The first one of these states that, under the transformation
(\ref{ferro14}) of the pion field, the vector $e_i(\pi)$ transforms according
to the adjoint representation
$D^i_{\;j}(g)=\delta^i_{\;j} +f^i_{\;jk}g^k+\ldots$ of G. Since the
action of the group is transitive on G/H, this property fully
determines the function $e_i(\pi)$ in terms of its values for $\pi=0$, i.e., in
terms of the magnetization.
The relation $(e)$ then specifies the rotation of
$c_a(\pi)$ and thus fixes the function itself up to a gradient. The reason
why $c_a(\pi)$ is not fully determined is that one may modify the Lagrangian
by a total derivative without changing the generating functional:
the operation $c_a(\pi)\rightarrow c_a(\pi)+\partial_a\omega(\pi)$
is equivalent to
${\cal
L}_{\eff}\rightarrow {\cal L}_{\eff}+\frac{\partial}{\partial t}\omega(\pi)$.
Except for this ambiguity, which is without physical significance,
the effective coupling constants of the new vertices
are fully determined by the order parameters $<\!0\!\mid\!J^0_i\!\mid\!0\!>$.

In the case of an abelian symmetry, $f^k_{\;ij}=0$, the relation $(d)$ shows
that
$e_i(\pi)$ is a constant and the condition $(e)$ then implies that $c_a(\pi)$
is a pure gradient and may thus be removed, $c_a(\pi)=0$. Accordingly, the term
${\cal L}_{\eff}^{(0,1)}$ takes the form $e_i(0)f^i_0$. Since this expression
does not involve the pion field, it leads a life of its own, exclusively
generating an expectation value for the charge densities. For abelian
symmetries, the equation of motion is, therefore of second order in the time
derivative, such that the dispersion law takes the form
$\omega\propto\,\mid\!\vec{k}\!\mid$.

{}From a methodical point of view,
the most remarkable property of the new couplings is
that the corresponding contribution to the effective Lagrangian in general
fails to be gauge invariant. Subjecting the fields $f_\mu^i$ and $\pi^a$ to
the infinitesimal gauge transformations (\ref{ferro5}) and (\ref{ferro14})
and using the relations $(d)$ and $(e)$, one finds that the
effective Lagrangian picks up a total derivative:
\be\label{ferro51}
\delta {\cal L}_{\eff}^{(0,1)}=\mbox{$\frac{\partial}{\partial
t}$}\{g^i\,[c_a(\pi)h^a_{\;i}(\pi) +e_i(\pi)]\}\fs\ee

Now, this may merely be due to a bad convention. If the expression in
square brackets is of the form $h^a_{\;i}(\pi)\partial_a \omega(\pi)$, it
suffices to modify the Lagrangian by a total derivative, to make it gauge
invariant (${\cal L}_{\eff}\rightarrow{\cal L}_{\eff}+\frac{\partial}{\partial
t}\omega(\pi)$). So, if the
function $e_i(\pi)$ is of the form
$e_i(\pi)=h^a_{\;i}(\pi)\bar{e}_a(\pi)$, with
$\bar{e}_a=-(c_a+\partial_a\omega)$, there is no problem with gauge invariance.
The condition, in particular, requires the vector $e_i(0)$ to be contained in
the subspace spanned by the Killing vectors at $\pi=0$.

Denote the
Lie algebras of G and H by {\bf G} and {\bf H}, respectively and set {\bf
G} = {\bf H} + {\bf K}. The Killing
vectors span the subspace {\bf K}. Hence, for the Lagrangian to be gauge
invariant,
the vector $e_i(0)$ must be contained in this subspace.

The Lie algebra {\bf G} transforms with the adjoint representation of G.
The corresponding representation of the subgroup H maps the two
subspaces {\bf H} and {\bf K} onto themselves; in particular, {\bf K} carries
a representation of H. Since
the order parameters $e_i(0)$ are invariant under H, this representation must
contain
a one-dimensional invariant subspace. Unless this is the case, the charge
densities can receive nonzero expectation values only if the Lagrangian
violates gauge invariance. For the ferromagnet, e.g., the magnetization
$e_i(0)$ belongs to {\bf H} rather than {\bf K} --- the corresponding
effective Lagrangian necessarily breaks gauge invariance.

\section{Ferromagnet}
\label{Ferromagnet}

I now discuss the case of the ferromagnet in some detail, i.e., consider the
groups
$\mbox{G}=\mbox{O(3)},\,\mbox{H}=\mbox{O(2)}$. The corresponding
structure constants are given by
$f^i_{\;jk}=\varepsilon_{ijk}$; there are
three conserved currents, $i=1,2,3$, and two pion fields, $a=1,2$.
In the
Heisenberg model, e.g., the magnetic moment of the lattice sites may be
represented
as $\mu \vec{s}_n$, where $\vec{s}_n$ is the spin of the site. In the
notation used here, the interaction with
a constant magnetic field, $\mu\sum_n \vec{s}_n\cdot \vec{H}$, corresponds to
the term $\int\!d^3\!xf_0^iJ^0_i$. The spin rotations are generated
by the total angular momentum, such that
$\sum_n\vec{s}_n=\int\!d^3\!x\vec{J}^0$. Accordingly, the time
components of the external field are related to the magnetic field by
$f_0^i=\mu H^i$.

It is convenient to use a covariant representation
for the pion
field, replacing the two variables $\pi^1,\pi^2$ by a three-dimensional unit
vector
$\vec{U}=(U^1,U^2,U^3)$, which transforms with
the vector representation of O(3). The nonlinear transformation law
(\ref{ferro14}) then takes the linear form $\delta
U^i=\varepsilon_{ijk}U^jg^k$. In this notation,
the term ${\cal L}_{\eff}^{(2,0)}$ is proportional to the square of the
covariant derivative of $\vec{U}$,
\be\label{ferro50}
{\cal L}_{\eff}^{(2,0)}=-\mbox{$\frac{1}{2}$}F^2
D_{\!r}U^iD_{\!r}U^i
\hspace{1em},\hspace{1em}D_{\!r}U^i=
\partial_rU^i+\varepsilon_{ijk}f^j_rU^k\co\ee
As mentioned above, symmetry determines the form of this part of the
Lagrangian, up to one
effective coupling constant, $F$. The corresponding explicit expressions for
the metric and for the Killing vectors are
$g_{ab}=F^2\partial_aU^i\partial_bU^i,\,h_{ai}=F^2
\varepsilon_{ijk}\partial_aU^jU^k$.

The analogous representation of the function
$e_i(\pi)$ immediately follows from the completeness
of the spherical
harmonics on the two-sphere: there is only one set of three
functions of the pion field transforming
according to the vector representation of O(3). Hence the vectors $e_i(\pi)$
and $U^i(\pi)$ are proportional to one another, $e_i=\Sigma\, U^i$.
The constant of proportionality $\Sigma$ is the magnitude of the
magnetization.

The expression for the function $c_a(\pi)$ is more complicated. Using the
completeness
relation for the Killing vectors, $\sum_ih_{ai}h_{bi}=F^2g_{ab}$, the condition
$(e)$ may be rewritten in the form
\be\label{ferro20}
\partial_ac_b-\partial_bc_a =-\Sigma\, \varepsilon_{ijk}\,\partial_aU^i\,
\partial_bU^j\,U^k\fs\ee
The right hand side is reminiscent of a topological invariant: up to a
factor of $4\pi\Sigma$, the integral
over the sphere is the Brouwer degree of the map $\vec{U}(\pi)$.

The
differential equation (\ref{ferro20}) may be integrated with the technique used
in the construction of the Wess-Zumino term. Consider a point $\pi$ on the
sphere and join it smoothly to $\pi=0$, along the path
$\sigma^a[\pi,\lambda],\,0\leq\lambda\leq 1$, with
$\sigma^a[\pi,0]=0,\;\sigma^a[\pi,1]=\pi^a$. Define the function $c_a(\pi)$
as the integral
\be\label{ferro21}
c_a(\pi)=\Sigma\int_0^1\!d\lambda \,
\varepsilon_{ijk}\,\partial_a U^i\,\partial_\lambda U^j \,U^k\co\ee
with $U^i=U^i(\sigma[\pi,\lambda])$. The vectors $\partial_a \vec{U}$ and
$\partial_\lambda \vec{U}$, which denote the derivatives with respect to
$\pi^a$ at constant $\lambda$ and vice versa, are orthogonal
to $\vec{U}$. Since the tangent plane only contains two linearly
independent directions, the quantity
$\varepsilon_{ijk}\,\partial_a U^i\,\partial_bU^j\,\partial_\lambda U^k$ is
equal to zero. Using this property, one readily checks that the function
defined in (\ref{ferro21}) indeed obeys the differential equation
(\ref{ferro20}). As noted above, any other solution differs from this one by an
irrelevant gradient.

Together with the contribution involving the external field, the Lagrangian
thus becomes
\be\label{ferro22}
{\cal L}_{\eff}^{(0,1)}=\Sigma\!\int_0^1\!d\lambda\,
\varepsilon_{ijk}\,\partial_0 U^i\,\partial_\lambda U^j \,U^k
+\Sigma f_0^i\,U^i\fs\ee
The form of the
path $\sigma[\pi,\lambda]$ affects the result only through a total derivative.
For the particular choice $U^i(\sigma[\pi,\lambda])=\lambda U^i(\pi)$,
$i=1,2$, the derivatives of the interpolating field may be
expressed in terms of those of $U^i(\pi)$ and
the integral may then be performed explicitly, with the result
\be\label{ferro22a} {\cal L}_{\eff}^{(0,1)}
=\Sigma\,(1+U^3)^{-1}(\partial_0U^1U^2-\partial_0U^2U^1)
+\Sigma f_0^i\,U^i\fs\ee
Visibly, the expression violates gauge invariance.

The corresponding equation of motion is obtained by evaluating the change in
the action generated
by a deformation of the pion field. Using the representation (\ref{ferro22})
for ${\cal L}^{(0,1)}_{\eff}$, the calculation yields
\be\label{ferro23} \Sigma\,\varepsilon_{ijk}U^j\dot{U}^k +\Sigma f_0^i
+F^2{\bf \Delta}U^i=\alpha\, U^i\co\ee
where ${\bf \Delta}=D_{\!r}D_{\!r}$ is the covariant Laplacian and $\alpha$ is
a Lagrange multiplier,
arising from the constraint $\delta\vec{U}\!\cdot\vec{U}=0$. The result may be
rewritten in the vectorial form
\be\label{ferro24}
\Sigma\, \dot{\vec{U}} +\Sigma\,\vec{f}_0 \hspace{-0.2em}\times
\hspace{-0.2em}\vec{U}
+F^2{\bf \Delta}\vec{U}\hspace{-0.3em}\times\hspace{-0.2em}\vec{U} =0\fs\ee
Indeed, this equation is known to describe the spin waves of a ferromagnet
--- it is referred to as the Landau-Lifshitz equation \cite{Anderson,Landau}.
The above discussion merely identifies a known model within the present
framework:
the Landau-Lifshitz equation is the equation of motion associated with the
leading terms in the derivative expansion of the general
effective Lagrangian, for G = O(3), H = O(2). The
Lagrangian contains a term related to the Brouwer degree of the map
$U^i(\pi)$;
the corresponding effective coupling constant is the expectation
value of the charge density.

The dispersion law of the spin waves may be worked out by
considering the fluctuations of the field in the vicinity
of the ground state $\vec{U}_0=\mbox{const.}$ Taking the magnetization
to point along the third axis, $\vec{U}_0=(0,0,1)$, the linearized equation
of motion only involves the particular combination
\be\label{ferro24a}
f^a=f_0^a+\gamma\,\varepsilon_{ab3} \partial_r f_r^b
\hspace{2em},\hspace{2em}\gamma\equiv\frac{F^2}{\Sigma\;}\ee
of external fields. Collecting the two transverse components of
$\vec{U}$ in a complex field $u=U^1+iU^2$, the equation of motion reduces to
\be\label{ferro25}- i\dot{u}-\gamma\!\bigtriangleup\! u=f\co\ee
with $f=f^1+if^2$.
So, the dispersion law of the magnons takes the form
\be\label{ferro25a}
\omega(\vec{k})=\gamma\vec{k}^{\,2}+O(k^4)\co\ee
where $\gamma$ is fixed by the two effective coupling constants $F$ and
$\Sigma$, according to (\ref{ferro24a}).

A constant magnetic field, $f_0^i=\mu H(0,0,1)$, explicitly breaks the
symmetry and generates a magnon "mass term", $\Sigma f_0^iU^i=
\Sigma\mu H (1-\frac{1}{2}u^\star u+\ldots)$, much like the quark masses
explicitly break the chiral symmetry of QCD, providing the pions with mass.
In the present case, the perturbation merely lifts the energy of all
lattice sites by $\mu H$, such that the dispersion law remains the same,
except for an overall shift,
$\omega(\vec{k})=\gamma\vec{k}^{\,2}+\mu H$.

\section{Correlation functions of a ferromagnet}
\label{Correlation}

The same coupling constants also determine the low energy behaviour of the
correlation functions
of the charge densities and currents. The corresponding two-point-functions are
given by the part of the generating functional which is quadratic in the
external field. At leading order of the low energy expansion, the generating
functional is the classical action of ${\cal
L}^{(0,1)}_{\eff}+{\cal L}^{(2,0)}_{\eff}$, evaluated at the solution of the
equation of motion. Since the functional collects
the {\it time-ordered} correlation functions, Feynman boundary
conditions are relevant:
the solution $U^i(x)$ is to contain only positive (negative)
frequencies as $t\rightarrow +\infty\, (-\infty)$. For the combination
$u=U^1+iU^2$, the solution is given by
\bea\label{ferro26}
u(x)&\hspace{-0.5em}=&\hspace{-0.5em}\int\!d^4\!y\,G(x-y)f(y)\co\\
G(x)&\hspace{-0.5em}=&\hspace{-0.5em}
\int\!\frac{d^3\!k\,d\omega}{(2\pi)^4}\,\frac{e^{i\vec{k}\vec{x}-i\omega
t}}{\;\gamma \vec{k}^{\,2}-\omega-i\epsilon}=
i\,\theta(t)\int\!\frac{d^3k}{(2\pi)^3}\,
e^{i\vec{k}\vec{x}-i\gamma\vec{k}^{\,2}t} \fs\nonumber\eea
Note that the Feynman solution is complex: the expression for the combination
$\bar{u}=U^1-iU^2$ does not coincide with the complex conjugate of the
solution $u$, but is determined by the complex conjugate of $f$, according to
$\bar{u}(x)=\int\!d^4\!y\,G(y-x)f^\star(y)$. The resulting
expression for the pion field is of the form
\be\label{ferro27}
U^a(x)=\int\!d^4\!y\,G_{ab}(x-y)\,f^b(y)\co\ee
where $G_{ab}(x)$ is the relevant Feynman propagator,
\be\label{ferro28}
G_{ab}(x)=G_{ba}(-x)=\mbox{$\frac{1}{2}$}\delta_{ab}\{G(x)+G(-x)\}
+\mbox{$\frac{1}{2}$}i\varepsilon_{ab3}\{G(x)-G(-x)\}\fs\ee
Note that $G_{ab}(x)$ describes the propagation of a single particle ---
although the effective theory contains two pion fields, there is only one
magnon of a given momentum. The propagator may be written in the form
\be\label{ferro29}
G_{ab}(x)=\varepsilon_a\varepsilon^\star_b\,G(x)+
\varepsilon_b\varepsilon^\star_a\,G(-x)\hspace{1em},\hspace{1em}
\varepsilon_a=\mbox{$\frac{1}{\sqrt{2}}$}(1,-i)
\co\ee
which explicitly shows the degeneracy of the propagation matrix.

Inserting the solution (\ref{ferro27})
in the expression for the action, one finally obtains
\be\label{ferro30}
\Gamma\{f\}=\int\!\!d^4\!x\,\Sigma\,f_0^3 -\mbox{$\frac{1}{2}$}F^2
\int\!\!d^4\!x\,f_s^af_s^a+
\mbox{$\frac{1}{2}$}\Sigma \!
\int\!\!d^4\!xd^4\!y\,f^a(x)\,G_{ab}(x-y)\,
f^b(y)+\ldots\ee
with $f^a=f_0^a+\gamma\,\varepsilon_{ab3} \partial_r f_r^b$. The term linear in
$f_0^3$ represents the one-point function,
$<\!0\!\mid\!J^0_i\!\mid\!0\!>=\delta^3_i\,\Sigma$. The coefficient
of the contribution which is quadratic in $f_0^a$ is the leading term
in the low energy expansion for the correlation function of the transverse
charge densities,
\be\label{ferro31}
<\!0\!\mid\!T\{J^0_a(x)J^0_b(0)\}\!\mid\!0\!>=(-i)\Sigma\,
G_{ab}(x)+\ldots\ee
The Fourier transform thereof contains a pole, whose
residue represents the square of the transition matrix element
$<\!0\!\mid\!J^0_a\!\mid\!\pi(\vec{k})\!>$
between the ground state and a magnon of momentum $\vec{k}$.
Using the nonrelativistic normalization
\be\label{ferro32}
<\!\pi(\vec{k}^\prime)\!\mid\!\pi(\vec{k})\!>
=(2\pi)^3\delta^3(\vec{k}^\prime-\vec{k})\co\ee
the result for the matrix element reads
\be\label{ferro33}
<\!0\!\mid\!J^0_a\!\mid\!\pi(\vec{k})\!> =
\varepsilon_a\,\sqrt{\Sigma} \fs\ee

Euclidean invariance requires the corresponding matrix element of the currents
to be proportional to the vector $\vec{k}$. Current conservation then shows
that the coefficient of proportionality is given by
\be\label{ferro34}
<\!0\!\mid\!J^s_a\!\mid\!\pi(\vec{k})\!> =
k^s\varepsilon_a\,\gamma\,\sqrt{\Sigma}=
k^s\varepsilon_a\,F^2/\sqrt{\Sigma} \fs\ee
The corresponding expression for the correlation function of the currents
is obtained by extracting the part of the generating functional which is
quadratic in $f_r^a(x)$. The result reads
\be\label{ferro31a}
<\!0\!\mid\!T\{J^r_a(x)J^s_b(0)\}\!\mid\!0\!>=iF^2\gamma\,\partial_r\partial_s
G_{ab}(x)+iF^2\delta_{rs}\delta_{ab}\delta^4(x)+\ldots\ee
The contact contribution $\propto \delta^4(x)$ is required by the Ward
identities; it arises from
the second term on the right hand side of (\ref{ferro30}).

With the above explicit form of the effective Lagrangian, it is a matter of
straightforward calculation to work out
magnon-magnon scattering amplitudes and to establish a low energy theorem
analogous to Weinberg's prediction for the scattering lengths of
$\pi\pi$ scattering \cite{Dashen}. Likewise, the expansion of the magnetization
in powers of the temperature may be evaluated by
repeating the analogous calculation for
the quark condensate \cite{Gerber}, where the expansion has been worked out
to order $T^6$. In that work, the explicit symmetry breaking due to the quark
masses is taken into account, indicating that the same methods also allow one
to study the perturbations generated by a weak, constant magnetic field.

Both the physics of magnon scattering and the structure of the
low temperature expansion for the magnetization is well understood since the
pioneering work of Dyson \cite{Dyson}. What a reanalysis of the same phenomena
by means of an effective Lagrangian may add is a
better understanding of the fact that many of the low energy properties of the
system are immediate consequences of the hidden symmetry, while the microscopic
structure of the system only manifests itself in the numerical values of
a few effective coupling constants. Also, the method may prove to be more
efficient, allowing one to carry the low energy expansion to higher orders.
Work on applications of the effective
Lagrangian constructed in the present paper is in progress \cite{Hofmann}.

\section{Antiferromagnet}
\label{Antiferromagnet}

Symmetry does not prevent the charge densities from picking up an expectation
value, but does not insure this to happen, either. The
antiferromagnet
is a well-known system where the expectation value of the spin density
vanishes.
The corresponding effective field theory is discussed
extensively in the recent literature \cite{Hasenfratz}. The work
goes beyond the leading terms of the low energy expansion and also
includes an analysis of the anisotropies generated by the
lattice. The present section does not add anything to what is
known
about antiferromagnetic systems. I merely wish to identify
these within the general framework of effective field theory and
to compare their low energy
structure with the one of the ferromagnet.

In the language of the effective Lagrangian,
antiferromagnets represent the special case where
the effective coupling constants $e_i(0)$ happen to be zero.
As discussed above, the conditions $(d)$ and $(e)$ then imply that
the functions $e_i(\pi)$ and $c_a(\pi)$ vanish altogether, such that
the derivative expansion of the effective Lagrangian only starts at second
order, with the contributions listed in (\ref{ferro11}).
The form of the functions $g_{ab}(\pi),h^a_{\;i}(\pi),k_{ik}(\pi),
\bar{g}_{ab}(\pi),\bar{h}^a_{\;i}(\pi),\bar{k}_{ik}(\pi)$ occurring therein
may be worked out along the same lines as before.
In the absence of the term ${\cal L}_{\eff}^{(0,1)}$, the standard power
counting used in particle physics, which
treats energies and momenta as quantities of the same algebraic
order, is more appropriate than the one introduced above. The
comparison of the
two differential equations (\ref{ferro8}), (\ref{ferro9}) then again leads
to a set of
conditions, which these functions need to satisfy for the effective Lagrangian
to give rise to a gauge invariant generating functional. In fact, the
conditions for $g_{ab}(\pi),h^a_{\;i}(\pi),k_{ik}(\pi)$ are identical with
those found previously: these quantities are subject to the conditions $(a),
(b)$ and $(c)$ of equation (\ref{ferro13}). Moreover, the barred
quantities must obey precisely the same constraints. The solution of these
conditions was discussed in section \ref{Symmetry}. As mentioned there, the
resulting expression for the effective Lagrangian is gauge invariant --- a
topological term only arises if the charge densities do pick up an
expectation value.

I again specialize to the groups G = O(3), H = O(2), where the
expectation
values of the charge densities are
given by $e_i(0)=\delta^3_i\,\Sigma$; the present discussion
thus concerns the special case $\Sigma =0$. As mentioned above,
the functions
$g_{ab}(\pi),h^a_{\;i}(\pi),k_{ik}(\pi)$ are also fixed up to a
constant. Since the barred quantities obey identical constraints, the same
is
true for these. The Lagrangian thus contains two copies of the same expression,
\be\label{ferro35}
{\cal L}_{\eff}^{(2)}=\mbox{$\frac{1}{2}$}
F^{\,2}_{\!1}D_0U^iD_0U^i- \mbox{$\frac{1}{2}$}F^{\,2}_{\!2}D_sU^iD_sU^i
\hspace{1em},\hspace{1em}D_\mu U^i=
\partial_\mu U^i+\varepsilon_{ijk}f^j_\mu U^k\fs
\ee
At leading order in the derivative expansion, the Lagrangian involves
two effective coupling constants,
$F_1$ and $F_2$. Except for the number of components of the vector
$\vec{U}$ and for the
magnitude of the constants $F_1$ and $F_2$,
the effective Lagrangian is the same as for QCD with two quark flavours or for
the Higgs sector of the standard model. There, the two coupling
constants
are related by the velocity of light, $F_2=cF_1$. This shows that, for the
antiferromagnetic systems under discussion here, euclidean
invariance implies Lorentz invariance, except that (i) the velocity of
light is to be replaced by $v\equiv F_2/F_1$ and (ii)
the statement only holds at leading order of the
low
energy expansion. In particular, the dispersion law corresponds to a massless
particle moving with velocity $v$,
\be\label{ferro36}
\omega(\vec{k})=v\!\!\mid\!\!\vec{k}\!\!\mid\!+\,O(k^2)\fs\ee

The known results of the low energy analysis for the strong interactions may
be taken over as they are, merely replacing the velocity of light by $v$ and
adapting the number of components of $\vec{U}$. There are now two magnons,
because the equation of motion for the effective field $\vec{U}$
happens to be of second order with respect to time. In the nonrelativistic
normalization of the states used in the preceding sections, the transition
matrix elements of the charge and current densities are given by
\be\label{ferro37}
<\!0\!\mid\!J^0_a\!\mid\!\pi^b(\vec{k})\!> =
i\,\delta_a^b\mid\!\vec{k}\!\mid \!F_2\,/\sqrt{2\omega}\hspace{1em},
\hspace{1em} <\!0\!\mid\!J^r_a\!\mid\!\pi^b(\vec{k})\!> =
i\,\delta_a^b\,k^r vF_2\,/\sqrt{2\omega}\fs\ee
In the case of the antiferromagnet, the transition elements of charge density
and current are of the same order in the momentum and tend to zero for
$\vec{k}\rightarrow 0$, while in the ferromagnetic case, they are of different
magnitude, the charge density generating transitions even at infinite
wavelength.

\section{Phonons in solids}

Historically, the phenomena associated with the propagation of sound were among
the very first to be analyzed in terms of an effective field theory.
For a solid, the re\-le\-vant effective fields are the components of the
vector $\vec{\xi}(x)=(\xi^1(x),\xi^2(x),\xi^3(x))$, which specifies the
displacement of the material
from the position in the ground
state.
The corresponding equation of motion follows from
the conservation of momentum,
\be\label{ferro200}
\partial_\mu\theta^{\mu r}(x)=\partial_0\theta^{0r}(x)+\partial_s\theta^{sr}(x)
=0\fs\ee
The quantity $\theta^{0r}$ is the momentum
per unit volume, while
$\theta^{rs}$ is the stress tensor describing the momentum flow per unit area
and time. To first order in the amplitude of the deformation, the
momentum density is proportional to the mass density
$\rho$ of the solid and to the velocity field,
\be\label{ferro201}
\theta^{0r}=\rho\,\dot{\xi}^r\fs\ee
For simplicity, I consider a cubic lattice.
Symmetry under reflections then implies that, to first order in the derivative
expansion, the stress tensor is invariant under rotations,
\be\label{ferro202}
\theta^{rs}=-\mu\, \xi^{rs}
-K\,\delta^{rs}\,\vec{\partial}\!\cdot\!\vec{\xi}\hspace{1em},\hspace{1em}
\xi^{rs}\equiv\partial_r\xi^s+\partial_s\xi^r-\mbox{$\frac{2}{3}$}\delta^{rs}\,
\vec{\partial}\!\cdot\!\vec{\xi}\fs\ee
The constants $\mu$ and $K$ are referred to as torsion and compression
modules, respectively \cite{Hydro}.

The
conservation law (\ref{ferro200}) shows that, at large wavelengths,
the sound waves of a solid obey the wave
equation \be\label{ferro203} \rho\;\ddot{\!\vec{\xi}}-\mu\bigtriangleup\!
\vec{\xi}-(K+\mbox{$\frac{1}{3}$}\mu)\,\vec{\partial}\,(\vec{
\partial}\!\cdot\!\vec{\xi}\,) =0\fs\ee The corresponding dispersion law is of
the form $\omega(\vec{k})=v\!\mid\!\vec{k}\!\mid\!+\,O(k^2)$. For
trans\-verse vibrations ($\vec{\xi}\perp\vec{k}$), the velocity of sound is
determined by the torsion module, $v_{\!\perp}=
\sqrt{\mu/\rho}$, while longitudinal waves propagate with
$v_{{\scriptscriptstyle \parallel}}=
\sqrt{(K+\frac{4}{3}\mu)/\rho}$.

The energy density also admits an expansion in powers of the effective field
and its derivatives. The leading contribution arises from the rest energy of
the material and is of first order in $\vec{\xi}$, while the energy of the wave
itself only shows up at second order.
The leading term is readily
obtained from the energy conservation law
\be\label{ferro202a}
\partial_0\theta^{00}+\partial_r\theta^{r0}=0\fs\ee
In view of the symmetry
$\theta^{\mu\nu}=\theta^{\hspace{0.05em}\nu\hspace{-0.05em}\mu}$, this yields
\be\label{ferro205}
\theta^{00}=-\rho\,(\vec{\partial}\!\cdot\!\vec{\xi}\,)\fs\ee
In the notation
used here, the energy density is given by $c^2\theta^{00}$, such that the
undeformed solid corresponds to $\theta^{00}=\rho$. The
expression (\ref{ferro205})
represents the change in the density generated by
the
deformation, $\rho \rightarrow\rho (1-\vec{\partial}\!\cdot\!\vec{\xi}\,)$.
Note that the contribution from the ground state itself is dropped, in
$\theta^{00}$ as well as in $\theta^{rs}$.

The sound waves may be viewed as Goldstone excitations of spontaneously
broken space-time symmetry: G is the Poincar\'{e} group and
H is the group of time translations (in the nonrelativistic domain
of interest here, G may equally well be identified with the Galilei group). The
elements of the quotient G/H are
parametrized by a rotation matrix $R$, a velocity $\vec{v}$ and a
space translation $\vec{a}$; the corresponding generators are the angular
momentum $\vec{J}$, the boost $\vec{K}$ and the momentum $\vec{P}$,
respectively. For spontaneously broken {\it internal} symmetries, the
effective theory involves as many pion fields as there are coordinates
in G/H. Accordingly, one might
expect that the effective field theory requires a
matrix
field $R(x)$ as well as two vector fields $\vec{v}(x),\vec{a}(x)$. The
standard analysis sketched above, however, only involves a single
vector field, $\vec{\xi}(x)$. Indeed, the fields $R(x)$ and $\vec{v}(x)$ are
redundant: the transformation law $\vec{x}\rightarrow
R\vec{x}+\vec{v}t+\vec{a}$ shows that space-time dependent translations
also cover boosts and rotations. The local form of the symmetry group G is
the set of general coordinate transformations; infinitesimally, these are
described by space-time dependent translations, $x^\mu\rightarrow
x^\mu+a^\mu(x)$. The spontaneously broken part thereof consists of the
translations in space. The
state $\exp\,i\,\vec{a}(x)\!\cdot\!\vec{P}\mid\!0\!>$
represents a deformed ground state, the point $\vec{x}$ being shifted into
$\vec{x}+\vec{a}(x)$. Hence the field $\vec{a}(x)$ coincides with the
effective field $\vec{\xi}(x)$ introduced above. One may thus view the phonons
of a solid as Goldstone bosons associated with spontaneously broken translation
invariance: at long wavelength,
$\mid\!\vec{k}\!\mid\rightarrow 0$, the frequency of the sound waves tends
to zero, because there is no restoring force for displacements of the
solid as a whole. The fact that the solid also breaks
invariance under
rotations and boosts does not give rise to additional Goldstone bosons.

The wave equation (\ref{ferro203})
is the equation of motion of the Lagrangian
\be\label{ferro204}
{\cal L}_{\eff}=\mbox{$\frac{1}{2}$}\rho\,\dot{\xi}^r\dot{\xi}^r
-\mbox{$\frac{1}{4}$}\mu\,\xi^{rs}\xi^{rs}
-\mbox{$\frac{1}{2}$}K(\vec{\partial}\!\cdot\!\vec{\xi}\,)^2 \co
\ee
which is of the same structure as the effective Lagrangian
relevant for the spin waves of an antiferromagnet. The
mass density plays the
role of the effective coupling constant $F_{\!1}^{\,2}$, while $\mu$ and $K$
are the analogues of $F_{\!2}^{\,2}$. In the present case (i) the
relevant symmetry group G is the abelian group formed by the
space-translations rather than the group O(3) considered in the preceding
section and (ii) the
symmetry is now fully broken, such that
there are three Goldstone bosons rather than two --- the subgroup H only
contains the unit element.

\section{Local form of the translation group}

As noted above, the relevant part of the space-time
symmetry which characterizes a solid is the group G of space-translations,
spontaneously broken to $\mbox{H}=\{e\}$. The generators of G are the three
components of the total momentum. Accordingly, the charge densities $J^0_i$
coincide with the components $\theta^{0i}$ of the energy-momentum tensor,
while the currents $J^r_i$ are represented by $\theta^{ri}$. Their
correlation functions may again be obtained by exposing the system
to an external field. In the present case, where the sources of interest are
the components of $\theta^{\mu\nu}(x)$, the relevant external field is
the gravitational field $g_{\mu\nu}(x)$. The Ward identities obeyed by the
correlation functions of the energy-momentum tensor are equivalent to the
statement that the generating functional is invariant under general coordinate
transformations.

The change in the Lagrangian due to the external field may be worked out
as follows.
Denote the metric of Minkowski space by $\eta_{\mu\nu}$ and set
$g_{\mu\nu}(x)=\eta_{\mu\nu}+f_{\mu\nu}(x)$.
Since the
energy-momentum tensor is the variational derivative of the action with
respect to the metric, the modification of the Lagrangian is
given
by ${\cal L}-\frac{1}{2}f_{\mu\nu}\theta^{\mu\nu} +O(f^2)$. The effective
Lagrangian picks up the analogous term linear in $f_{\mu\nu}$, involving the
above representations of the
energy-momentum
tensor within the effective theory. In the presence of an external
gravitational field, the effective Lagrangian thus becomes
\bea\label{ferro206}
{\cal L}_{\eff}=&\hspace{-0.5em}
\mbox{$\frac{1}{2}$}&\hspace{-0.8em}\rho\,\dot{\xi}^r\dot{\xi}^r
-\mbox{$\frac{1}{4}$}\mu\,\xi^{rs}\xi^{rs}
-\mbox{$\frac{1}{2}$}K(\vec{\partial}\!\cdot\!\vec{\xi}\,)^2\\
&\hspace{-0.5em}+&\hspace{-0.7em}\mbox{$\frac{1}{2}$}
f_{00}\,\rho\,\vec{\partial}\!\cdot\!\vec{\xi} -f_{0r}\,\rho\,\dot{\xi}^r
+\mbox{$\frac{1}{2}$}f_{rs}\,\mu\,\xi^{rs}
+\mbox{$\frac{1}{2}$}f_{rr}\,K\,\vec{\partial}\!\cdot\!\vec{\xi}
+\ldots\nonumber\eea
The conservation laws insure that the corresponding action is invariant under
the transformation
\be\label{ferro207}
f_{\mu\nu}\rightarrow f_{\mu\nu}+\partial_\mu a_\nu
+\partial_\nu a_\mu\ee
of the gravitational field --- the linearized form of a
general coordinate transformation on Minkowski space amounts to an
abelian gauge transformation. If the field $\vec{\xi}(x)$ solves the wave
equation for $f_{\mu\nu}(x)$, then the solution belonging to the transformed
gravitational field is given by $\vec{\xi}(x)+\vec{a}(x)$. Under a gauge
transformation of the external field, the effective field thus transforms
according to
\be\label{ferro208}
\vec{\xi}\rightarrow \vec{\xi}+ \vec{a}\fs\ee
Although the time component $a^0$ of
the coordinate transformation
changes the Lagrangian by a total derivative, it does not affect the solution
at all. The essential part of the symmetry is contained in the spacial
components $\vec{a}$, which represent the gauge transformations associated
with the symmetry group G. Under these, the quantities
$\dot{\xi}^r-f_{0r}$ and $\partial_r\xi^s+\partial_s\xi^r-f_{rs}$ are
gauge invariant. Completing the squares in
(\ref{ferro206}), one thus arrives at a gauge invariant effective Lagrangian,
\bea\label{ferro210}
{\cal L}_{\eff}&\hspace{-0.5em}=&\hspace{-0.5em}
\mbox{$\frac{1}{2}$}\rho\, D_0\xi^rD_0\xi^r
-\mbox{$\frac{1}{4}$}\mu\, \Xi^{rs}\Xi^{rs}
-\mbox{$\frac{1}{2}$}K(D_r\xi^r)^2+
\mbox{$\frac{1}{2}$}f_{00}\,\rho\, D_r\xi^r+\ldots\\
D_0\xi^r&\hspace{-0.5em}=&\hspace{-0.5em}
\dot{\xi}^r-f_{0r}\hspace{1em},\hspace{1em}D_r\xi^r=\partial_r\xi^r
-\mbox{$\frac{1}{2}$}f_{rr}\hspace{1em},\hspace{1em}
\Xi^{rs}=\xi^{rs}-f_{rs}+\mbox{$\frac{1}{3}$}\delta^{rs}f_{tt}
\fs\nonumber \eea

We could just as well have applied the
machinery of the preceding sections to the case of the translation group.
Introducing external fields couplied to the charge densities $\theta^{0i}$ and
currents $\theta^{ri}$ and imposing gauge invariance, the result would
have been the same (except for the additional term involving the
external field $f_{00}$, which is not related to the charge densities and
currents of the group G). The
point is that the Lagrangian describing the phonons of a solid emerges from the
above general discussion as the special case which corresponds to the abelian
symmetry of the translation group. In particular, sound waves illustrate
the remark made in section \ref{Symmetry}, according
to which the Goldstone bosons generated by the
spontaneous breakdown of an abelian group obey a dispersion law of the form
$\omega\propto\mid\!\vec{k}\!\mid$.
The transition matrix elements of the charge densities and currents between
the ground state and a Goldstone boson may also
be calculated in the same manner as before, with the result
\bea\label{ferro211}
<\!0\!\mid\!\theta^{00}\!\mid\!\pi(\vec{k})\!>&\hspace{-0.5em}=&\hspace{-0.5em}
-i\sqrt{\rho}\;\vec{k}\!\cdot\!\vec{\varepsilon}
/\sqrt{2\omega}\\
<\!0\!\mid\!\theta^{0r}\!\mid\!\pi(\vec{k})\!>
&\hspace{-0.5em}=&\hspace{-0.5em}-i\sqrt{\rho}\;\omega\,
\varepsilon^r/\sqrt{2\omega}\no
<\!0\!\mid\!\theta^{rs}\!\mid\!\pi(\vec{k})\!>
&\hspace{-0.5em}=&\hspace{-0.5em}-i\sqrt{\rho}\,\{
v_{\! \perp}^2(k^r\varepsilon^s+k^s\varepsilon^r-
2\delta^{rs}\vec{k}\!\cdot\!\vec{\varepsilon})
+v_{{\scriptscriptstyle
\parallel}}^2\delta^{rs}\vec{k}\!\cdot\!\vec{\varepsilon}\}
/\sqrt{2\omega}\co\nonumber\eea
where $\vec{\varepsilon}$ is the polarization vector of the phonon.

Despite these evident similarities with the spontaneously broken {\it internal}
symmetries discussed
in the preceding sections, the fact that the translation group
acts on space-time gives rise to some peculiarities. I add two
remarks regarding
the difference between phonons and Goldstone bosons of an internal symmetry.

The first point concerns the transformation properties of the
generators under space rotations. While the charges considered in the
preceding sections were assumed to be invariant, the
generators of the translations transform with the vector
representation of the rotation group. Euclidean
invariance then prevents the charge densities from acquiring expectation
values, $<\!0\!\mid\!\theta^{0r}\!\mid\!0\!>=0$, while those of
the
currents may be different from zero,
$<\!0\!\mid\!\theta^{rs}\!\mid\!0\!>=\delta^{rs}p$ ($p$ is the pressure in
the ground state). Apart from this modification, the general discussion of
section \ref{Symmetry}, however, applies. As pointed out there,
the order parameters associated with the charges of an abelian
group lead a life of their
own and do not manifest themselves in the dynamics of the Goldstone bosons.
Indeed, in the above equations, the contribution to the energy-momentum
tensor from the ground state were simply dropped.

The second remark is more significant.
The form of the Ward identities is controlled by the {\it local} version of the
symmetry group. The local form of the translation group is
the set of general coordinate transformations and is not abelian --- it reduces
to the
set of abelian gauge transformations (\ref{ferro207}) only at the linearized
level.
The intrinsic difference between the global and the local structure of the
group shows up in the commutation rules: while the
generators $P^r$ of the translation
group commute among themselves, they do not commute with the corresponding
charge densities and currents, but obey a commutation rule of the form $[P^r,
\theta^{\mu \nu}(x)]=i\hbar\partial_r \theta^{\mu \nu}(x)$.
The phenomenon is related to the fact that an abelian group admits different
local versions. For the
deformations of a solid, the one which matters is the group of coordinate
transformations, while for the
groups associated with the U(1)-charges of particle physics, the local form
relevant for the Ward identities is the set of abelian gauge
transformations. The full effective Lagrangian describing the
deformations of a solid is gauge invariant under the transformation
(\ref{ferro207}) only at the linearized level considered above. When imposing
the symmetry on the higher order terms, the expansion of the coordinate
transformation is needed
to higher accuracy, such that the transformation laws of the fields
$f_{\mu\nu}$ and $\xi^r$ then involve additional terms.

\section{Phonons in fluids and gases}

Finally, I briefly
comment on sound waves in fluids or gases. Since the corresponding ground state
is invariant under rotations as well as translations, the
generators $\vec{J}$ and $\vec{P}$ now belong to the subgroup H.
The spontaneously broken part of the space-time symmetry, G/H, is generated by
the boost operators $\vec{K}$. Accordingly, the effective field is the field
associated with space-time dependent boosts, $\vec{v}=\vec{v}(x)$. To lowest
order in this field, the leading terms in the derivative expansion of the
energy-momentum tensor now take the form
\be\label{ferro250}
\theta^{00}=\rho\hspace{1em},\hspace{1em}\theta^{0r}=\rho\,v^r
\hspace{1em},\hspace{1em}\theta^{rs}=p\,\delta^{rs}\ee
and the conservation laws for energy and momentum become
\be\label{ferro251}
\dot{\rho}+\vec{\partial}\!\cdot\!(\rho\,\vec{v}\,)=0
\hspace{1em},\hspace{1em}
(\rho\,\vec{v}\,)\,\dot{ }+\vec{\partial}\,p=0\fs\ee
In general, the local configuration of the system
depends on several variables, which, in principle are independent of one
another: in addition to the temperature, the
chemical potentials of the various particle species also need to be specified.
To the extent that the sound waves represent adiabatic deformations, the
change in the pressure is, however, determined by the one in the
density, $\delta p=\kappa\,\delta \rho$. The coefficient of proportionality is
the adiabatic compression module per unit mass, $\kappa=(
\partial p/\partial \rho)_s$ (a detailed discussion, in particular also of the
adiabatic
approximation, may be found in ref.
\cite{Hydro}). Eliminating $\dot{p}$ in favour of $\dot{\rho}$ and retaining
only terms linear in the velocity field, the time derivative of the
momentum conservation law may be rewritten as
\be\label{ferro252}
\ddot{\vec{v}}-\kappa\,\vec{\partial}\,(\vec{\partial}\!\cdot\!\vec{v}\,)=0\fs
\ee
The phonons thus obey a wave equation which is
similar to the one valid in solids ($\vec{v}\leftrightarrow\dot{\!\vec{\xi}}$).
The term proportional
to $\kappa$ is the analogue of the one involving the compression module,
$K=\rho\,\kappa$. A torsion term, on the other hand, does not occur here:
in fluids or gases,
torsion does not generate stress. A divergence free
velocity field
obeys $\,\ddot{\!\vec{v}}\,=0$, indicating that transverse modes do not
oscillate. According to equation (\ref{ferro252}), layers
perpendicular to the wave vector $\vec{k}$ glide along one another without
transfer of energy or momentum.

In reality, the energy contained in the transverse modes dissipates. The
attenuation rate is determined by the viscosity of the material, which
manifests itself in the stress tensor, at the next order of the derivative
expansion \cite{Hydro},
\be\label{ferro212}
\theta^{rs}=
\delta^{rs}\,p-\eta\{\partial_r\vec{v}^{\,s}
+\partial_s\vec{v}^{\,r}
-\mbox{$\frac{2}{3}$}\delta^{rs}\,\vec{\partial}\!\cdot\!
\vec{v}\,\} -\zeta\delta^{rs}\vec{\partial}\!\cdot\!\vec{v}\fs\ee
Instead of a wave equation, the transverse modes obey a diffusion
equation,
\be\label{ferro213}
\rho\;\dot{\!\vec{v}}=\eta\bigtriangleup\!\vec{v}\fs\ee
One may thus
conclude that, in the case of fluids or gases, there is only one Goldstone
particle. The two other degrees of freedom of the
group G/H are dissipative and do not propagate like particles with real
momenta and energies.
Instead, the corresponding
"dispersion law" corresponds to a pole in the complex plane, occurring at
$\omega(\vec{k})=-i\vec{k}^{\,2}\eta/\rho$.

For the effective Lagrangian method, this, unfortunately, is the end.
In the presence of phenomenological dissipative forces, the equation of motion
cannot be formulated in terms of a Lagrangian. This does not mean that
effective
field theory is unable to cope with the motion of fluids or gases --- quite to
the contrary, the Navier-Stokes equations describe this motion perfectly well.
They do represent an effective field theory, for which the velocity field
$\vec{v}(x)$ is the relevant dynamical variable.
That theory, however, cannot be represented in terms of an effective
Lagrangian. The systematic expansion in powers of the
derivatives provided by the effective Lagrangian method is not
available here. In this expansion, the contributions arising at higher
orders, from simultaneous exchange of several
Goldstone bosons, are accounted for by the loop graphs, i.e., by
the quantum fluctuations of the effective field. If
the effective field theory does not admit a Lagrangian formulation, it is
entirely unclear
how to set up the corresponding quantum theory. Presumably, in the
presence of phenomenological dissipative terms, it is impossible to
extend
the low energy analysis beyond leading order.

\section{Summary and conclusion}

$\hspace{1.5em}$1. The paper deals with
the effective field theory relevant for
the low energy analysis of spontaneously broken symmetries in the
nonrelativistic domain. The discussion applies to any system for which the only
excitations
without an energy gap are the Goldstone modes.

2. The analysis is based on the Ward identities obeyed by the correlation
functions of the charge densities and currents.
The discussion assumes
that
the Ward identities are anomaly free and exploits the fact that the
generating functional is then invariant under gauge transformations, i.e.,
under a local form of the symmetry group.

3. The number of effective fields
needed turns out to be universal. Denoting the symmetry groups of the
Hamiltonian and of the ground state by G and H, respectively,
the number of effective {\it fields} required to describe the properties of the
system
for large wavelengths is given by $\mbox{dim(G)} - \mbox{dim(H)}$. While, for
relativistically
invariant theories, the number of Goldstone {\it particles} coincides with the
number of effective fields, this is not in general the case for
nonrelativistic systems,
where the above number only represents an upper bound.

4. Nonrelativistic kinematics does not prevent the
generators of the group
from having expectation values in the ground state, representing order
parameters of the spontaneously broken symmetry.
The
main result of the present paper is the statement that
the phenomenon manifests itself through a term in the effective Lagrangian,
which is of topological nature and does not occur in the effective
field theories relevant for particle physics. The relevant term, in particular,
violates
gauge invariance of the effective Lagrangian.

5. The form of the leading
contributions in the derivative expansion of the general Lagrangian is
discussed in detail. In the case of
G = O(3), H = O(2), the two Goldstone fields may be described in terms of
a three-component vector $U^i(x)$ of unit length, $U^iU^i=1$. The derivative
expansion of the effective Lagrangian then starts with
\be\label{ferro260} {\cal L}_{\eff}=\Sigma\!\int_0^1\!d\lambda\,
\varepsilon_{ijk}\,\partial_0 U^i\,\partial_\lambda U^j \,U^k
+\mbox{$\frac{1}{2}$}
F^{\,2}_{\!1}\partial_0U^i\partial_0U^i-
\mbox{$\frac{1}{2}$}F^{\,2}_{\!2}\partial_rU^i\partial_rU^i+\ldots\ee
The first term is the topological object mentioned above.
The
corresponding effective
coupling constant $\Sigma$ is the order parameter associated with the charge
densities.
In the case of a magnet, $\Sigma$ is the magnetization of the ground state. The
other two effective coupling constants, $F_1,\,F_2$, are determined by the
one-particle matrix elements of the charge densities and currents.

6. The above expression for the effective Lagrangian implies that the
dispersion law of the Goldstone bosons is of the form \be\label{ferro261}
\Sigma \,\omega +F_{\!1}^{\,2}\omega^2
-F_{\!2}^{\,2}\vec{k}^{\,2}+\ldots=0\fs\ee (i) If the charge density acquires a
nonzero expectation value --- as it is the case with the ferromagnet ---
the first term is different from zero.
At low frequencies, it then dominates over the second,
such that
the dispersion law is quadratic in $\vec{k}$,
\be\label{ferro262}
\omega(\vec{k})=(F_{\!2}^{\,2}/\Sigma)\,\vec{k}^{\,2}+
\ldots\hspace{2em}\Sigma\neq 0\fs\ee
The corresponding wave equation takes the
form of
a Schr\"{o}dinger equation. The wave function is complex and incorporates
both of the two real Goldstone fields. The spectrum only contains
one Goldstone particle of a given momentum.\\
(ii) The antiferromagnet corresponds to the case where the charge density does
not acquire an expectation value. The
dispersion law then takes the form
\be\label{ferro263}
\omega(\vec{k})=(F_2/F_1)\!\mid\!\vec{k}\!\mid\!+\ldots\hspace{2em}
\Sigma=0\fs\ee
In this case, the wave equation is of second order in the time derivative,
such that there are two Goldstone particles.

7. The phonons of a solid represent a peculiar case, as they are associated
with a spontaneously broken space symmetry, translation invariance.
The relevant gauge group is the set of
coordinate transformations. Accordingly, the Ward identities for the
correlation functions of the energy-momentum tensor play a central role in the
corresponding effective theory.

8. While the effective Lagrangian method is
perfectly suited for the low energy analysis of the deformations of a solid,
the
method fails for fluids or gases. There, the low energy behaviour of two of the
three effective fields is dominated by
dissipative forces, which cannot be described in terms of a
Lagrangian.

\subsection*{Acknowledgement}
I thank U.
Wiese for clarifying discussions at an early phase of this work
and I am indebted to J. Gasser,
P. Hasenfratz, C. Hofmann, V. V. Lebedev, P. Minkowski, F. Niedermayer, A.
V. Smilga and U. W\"{u}rgler for comments, encouragement and help with the
literature. 

\begin{thebibliography}{99}

\bibitem{Anderson}
P. W. Anderson, {\it Basic notions of condensed matter physics} (Benjamin,
Menlo
Park, 1984);\\
H. Kleinert, {\it Gauge fields in condensed matter} (Worlds Scientific,
Singapore, 1989).

\bibitem{Dashen} S. Weinberg, Phys. Rev. Lett. {\bf 18} (1967) 188, 507; Phys.
Rev. {\bf 166} (1968) 1568;\\
R. Dashen, Phys. Rev. {\bf 183} (1969) 1245;\\
R. Dashen and M. Weinstein, Phys. Rev. {\bf 183} (1969) 1291.

\bibitem{Callan} S. Coleman, J. Wess and B. Zumino, Phys. Rev. {\bf 177} (1969)
2239;\\
C. Callan, S. Coleman, J. Wess and B. Zumino, Phys. Rev. {\bf 177} (1969)
2247.

\bibitem{Pagels} L.-F. Li and H. Pagels, Phys. Rev. Lett. {\bf 26} (1971)
1204;\\
H. Pagels, Phys. Rep. C {\bf 16} (1975) 219.

\bibitem{Weinberg1979} S. Weinberg, Physica {\bf A96} (1979) 327.

\bibitem{GL} J. Gasser and H. Leutwyler, Ann. Phys. (N.Y.) {\bf 158}
(1984) 142; Nucl. Phys. {\bf B 250} (1985) 465.

\bibitem{Mermin} N. D. Mermin and H. Wagner, Phys. Rev. Lett. {\bf 17} (1966)
1133.

\bibitem{Goldstone} J. Goldstone, Nuovo Cim. {\bf 19} (1961) 154.

\bibitem{Guralnik} A detailed discussion of the Goldstone theorem, in
particular also in the context of nonrelativistic broken symmetries is given
in\\
G. S. Guralnik, C. R. Hagen and T. W. B. Kibble, in
{\it Advances in particle physics}, Vol. 2, p. 567, ed. R. L. Cool and R. E.
Marshak
(Wiley, New York, 1968).

\bibitem{Foundations} H. Leutwyler, {\it On the foundations of chiral
perturbation
theory}, preprint Universit\"{a}t Bern, BUTP-93/24.

\bibitem{Landau} L. D. Landau and E. M. Lifshitz, {\it Course of Theoretical
Physics},
Vol. 9, {\it Statistical Physics}, Part 2, by
E. M. Lifshitz and L. P. Pitajewski (Pergamon, London, 1981).

\bibitem{Gerber} P. Gerber and H. Leutwyler, Nucl. Phys. {\bf B 321} (1989)
387.

\bibitem{Dyson} F. J. Dyson, Phys. Rev. {\bf 102} (1956) 1217, 1230.

\bibitem{Hofmann} C. Hofmann, PhD thesis, in preparation.

\bibitem{Hasenfratz}
H. Neuberger and T. Ziman, Phys. Rev. {\bf B 39} (1989) 2608;\\
P. Hasenfratz and F. Niedermayer, Phys. Lett. {\bf B 268} (1991) 231;
Z. Phys. {\bf B 92} (1993) 91;\\
U. J. Wiese and H. P. Ying, {\it Determination of the low energy parameters of
the
2D Heisenberg antiferromagnet}, preprint Universit\"{a}t Bern, BUTP-92/50.

\bibitem{Hydro}  L. D. Landau and E. M. Lifshitz, {\it Course of Theoretical
Physics}, Vol. 7, {\it Elasticity Theory} (Pergamon, London, 1959).

\end{thebibliography}
\end{document}